\documentclass[12pt,preprint]{aastex}

\def\lta{{\>\rlap{\raise2pt\hbox{$<$}}\lower3pt\hbox{$\sim$}\>}}
\def\gta{{\>\rlap{\raise2pt\hbox{$>$}}\lower3pt\hbox{$\sim$}\>}}

\def\R23{\mbox{$\rm R_{23}$}}

\def\am{\mbox{\AA\,}}

\def\arcsec{\hbox{$^{\prime\prime}$}}

\def\kmsmpc{km s$^{-1}$ Mpc$^{-1}$}

\def\Hb{\mbox{${\rm H}{\beta}$}}
\def\Ha{\mbox{${\rm H}{\alpha}$}}
\def\OIIIa{\mbox{${\rm [O\,III]\,}{\lambda\,5007}$}}

\def\OII{\mbox{${\rm [O\,II]\,}{\lambda\,3727}$}}

\def\NII{\mbox{${\rm [N\,II]\,}{\lambda\,6584}$}}

\shorttitle{Metallicities of $z\sim1.4$ galaxies}
\shortauthors{Maier, C. et al.}

\begin{document}


\title{Oxygen Gas Abundances at $z\sim 1.4$: Implications
  for the Chemical Evolution History of Galaxies\footnotemark[0]}

\author{C. Maier\altaffilmark{1}}
\email{chmaier@phys.ethz.ch}


\author{S.J. Lilly\altaffilmark{1}, C. M. Carollo\altaffilmark{1},
K. Meisenheimer\altaffilmark{2}, H. Hippelein\altaffilmark{2},
and A.  Stockton\altaffilmark{3}}

\footnotetext{Based on observations
  obtained at the ESO VLT, Paranal, Chile; ESO program 074.B-0122}
\altaffiltext{1}{Department of Physics, Swiss Federal Institute of Technology (ETH Z\"urich), ETH H\"onggerberg, CH-8093, Z\"urich, Switzerland}
\altaffiltext{2}{Max--Planck--Institut f\"ur Astronomie, K\"onigstuhl 17, 
           D-69117 Heidelberg, Germany}
\altaffiltext{3}{Institute for Astronomy, University of Hawaii, 2680
  Woodlawn Drive, Honolulu, HI 96822}


\begin{abstract}
  The $1<z<2$ redshift window hosts the peak of the star formation and
  metal production rates. Studies of the metal content of the star forming
  galaxies at these epochs are however sparse. We report VLT-ISAAC
  near-infrared spectroscopy for a sample of five [OII]-selected,
  $M_{B,AB} \la -21.5$, $z\sim1.4$ galaxies, by which we  measured \Hb\, and \OIIIa\,
  emission line fluxes from J-band spectra, and \Ha\, line fluxes plus upper
  limits for \NII\, fluxes from H-band spectra.  The $z\sim1.4$
  galaxies are characterized by
  the high [OIII]/[OII] line ratios, low extinction and low metallicity that
  are typical of lower luminosity CADIS galaxies at $0.4<z<0.7$, and of
  more luminous Lyman Break
  Galaxies at $z\sim 3$, but not seen in  CFRS galaxies at $0.4 \la z
  \la 0.9$. 
This type of spectrum (e.g., high \OIIIa/\OII\,) is seen in
progressively more luminous galaxies as the redshift increases.
These spectra are caused by a combination of high ionisation parameter
$q$ and lower [O/H].
P\'egase2 chemical evolution models are used to relate the observed
    metallicities and luminosities of $z\sim1.4$ galaxies to galaxy samples at
    lower and higher redshift.  Not surpringsingly, we see a relationship
    between redshift and inferred chemical age.  We suppose that the
    metal-enriched reservoirs of star forming gas that we are probing at
    intermediate redshifts are being mostly consumed to build up both the disk
    and the bulge components of spiral galaxies. Finally, our analysis
    of the metallicity-luminosity relation at $0 \la z \la 1.5$
    suggests that the period of rapid chemical evolution may take place
    progressively in lower mass systems as the universe ages.
These results are consistent with a ``downsizing'' type picture in
  the sense that particular signatures (e.g., high [OIII]/[OII] or low [O/H]) are seen in
  progressively more luminous (massive) systems at higher redshifts.
\end{abstract}

\keywords{
galaxies: abundances,
galaxies: evolution,
galaxies: high redshift
}


\section{Introduction}

~~~ Gas metallicities are a particularly important diagnostic of galaxy
evolution: at any given epoch, in any given galaxy population, they trace the
amount of metals which have been produced and assembled by its previous
stellar generations, as well as the metals which will be locked in its future
generations of stars.

~~~ Efforts to determine the gas metallicity of star forming galaxies as a
function of cosmic time  have returned oxygen ([O/H]) gas metallicities
for relatively large sample of galaxies up to $z \sim 1$ and a handful  above $z
\sim 2$ \citep{
  calilly01, lilly03, kob03,maier04,pettini02, maier05}. Most of
these studies have relied on rest-frame optical lines, and in particular on
the $\rm R_{23}= ({\rm [O\,II]\,}{\lambda\,3727} + {\rm
  [O\,III]\,}{\lambda\lambda\,4959,5007})/ {\rm H}{\beta}$ metallicity
indicator \citep{pagel79}. While relatively easy to measure, $R_{23}$ is
however degenerate with metallicity (as low values of $R_{23}$ may correspond
to very low or high metallicities), and  affected by dust extinction.
Both problems are solved if the ${\rm H}{\alpha}$ and ${\rm
  [N\,II]\,}{\lambda\,6584}$ line fluxes are also available, since the ${\rm
  H}{\alpha}/{\rm H}{\beta}$ line ratio provides an estimate for the dust
extinction, and the ${\rm [N\,II]\,}{\lambda\,6584}$/${\rm H}{\alpha}$ line
ratio breaks the degeneracy in $\rm R_{23}$.

~~~ The reason for which most quoted studies have not used the ${\rm
  H}{\alpha}$ and ${\rm [N\,II]\,}{\lambda\,6584}$ lines is a practical one.
For $z\gta 0.5$  these two lines are shifted in the
near-infrared (NIR), and are thus much more challenging to measure than the
lines which appear within the optical window. Nonetheless, these two lines
turn out to be important to the study of galaxy properties at high-z. In Paper
III we used VLT-ISAAC and Keck-NIRSPEC NIR spectroscopy to measure ${\rm
  H}{\alpha}$ and ${\rm [N\,II]\,}{\lambda\,6584}$ emission line fluxes -- and thus
individual extinction values and reliable metallicities -- for 30
$0.47<z<0.92$ galaxies extracted from the Canada France Redshift Survey 
\citep[CFRS]{lilly95}. We measured a large $\sim 3$\,mag scatter in extinction values for the CFRS galaxies,
which demonstrated the importance of individual extinction corrections in
order to avoid errors in the determination of star forming rates and
  metallicities of high $z$ galaxies.

~~~ The next step is to extend a similar study of the star forming gas
metallicity to galaxies at higher redshifts.  The rather unexplored $1<z<2$
redshift regime is one of particular importance in the history of the
universe: there, the star formation  and metal production rates for the
universe as a whole, as measured by the the integrated luminosity density in
the ultraviolet and far-infrared, appear to peak, i.e., are a factor of about 6
higher relative to the local value \citep[see, e.g.,][]{lilly96,madau96,charelbaz01,somerv01,pergonz05}.
  Furthermore, this is the redshift regime where the
galaxy population clearly undergoes a transition in properties: it is beyond
$z \sim 1$ that luminous ultraviolet star-forming galaxies with ``unobscured'' star-formation
rates (SFRs) $\dot M > 10$ M$_{\odot}$yr$^{-1}$, e.g., galaxies with
$L_{[OII]3727} > 10^{42}$ erg s$^{-1}$ \citep{cowie95}, appear in optically-selected galaxy
samples.  Such galaxies are not detected below $z \sim 0.8-1.0$ \citep[see,
e.g.,][]{cowie95}.

~~~ Due to the paucity in observable windows of spectral diagnostics for
galaxies in the $1<z<2$ redshift range,  the gas metallicities (and
more generally the physical properties) of $1<z<2$ galaxies are still mostly
unconstrained.  Luckily, for the specific redshift $z\sim 1.4$, all five
emission lines which are key to determine the gas metallicity and the
extinction, i.e., \OII, \Hb, \OIIIa, \Ha, and [NII], are all observable within
several  (near-infrared) atmospheric windows (there is also another such
window at $z\sim2.3$). 
At the extreme redshift limit of ground-based optical surveys \OII\,
 can be detected up to $z\sim1.5$ in the 9200 {\AA} atmospheric window,
 e.g. in the CFRS survey. 
 Furthermore, other surveys have exploited explicitly the 9200 {\AA}
 atmospheric window:
 the CADIS survey \citep[see e.g.,][]{maier02} and other
searches  for $z \sim 6.6$ Lyman $\alpha$
emitters \citep{cramlilly99,tran04} have produced, as  foreground
interlopers, several \OII-selected $z\sim 1.4$
galaxies.  The $z \sim 1.4$ epoch lies roughly midway in cosmic time between
the $0.5 < z < 1.0$ redshift range probed, e.g., by the CADIS
 \citep{maier04} and CFRS
galaxies (Paper\,III), and the $z \sim 3$ epoch characterized by the LBGs \citep[for a
handful of which oxygen abundances have been measured by][]{pettini01}.

~~~ In this paper we present oxygen gas abundances for a sample of five
galaxies at $z\sim 1.4$ selected by their \OII\, emission.  The paper is structured as follows: In Sect. 2 we
describe the VLT-ISAAC observations of our sample galaxies, and present the
lower  redshift samples that we use for comparison in our
discussion.  In Sect. 3 we present the derived emission line fluxes, 
metallicities  and extinction values derived (as in Paper\,III)
using a representation of the \citet{kewdop02} models. In Sect. 4 we present a grid of
P\'egase2 chemical evolution models, and discuss those which are compatible
with the observations.  We then compare our metallicity measurements at
$z\sim1.4$ with similar measurements at other redshifts, and with the
P\'egase2 models, and discuss the picture for the chemical evolution of
galaxies that emerges from our study. Finally, in Sect. 5 we present
our conclusions.
A {\sl concordance}-cosmology with $\rm{H}_{0}=70$ \kmsmpc,
$\Omega_{0}=0.3$, $\Omega_{\Lambda}=0.7$ is used throughout this
paper.  Our measurements of [O/H] do not depend on the solar [O/H]. However, when linking the
oxygen abundance with the metallicity calculated in   P\'egase2 models
we use the solar value
$12+\rm{log}(\rm{O/H})=8.87$ from \citet{grevesse}.  Note that {\sl
  metallicity} and {\sl abundance} will be taken to denote {\it oxygen abundance}
throughout this paper, unless otherwise specified.

%
%

\section{The data}

\subsection{Sample, observations and data reduction}

 ~~~ The sample galaxies were selected as \OII\,- emitters  at
 $z \sim 1.4$ from three sources:  the CADIS survey \citep[see, e.g.,][]{maier02,maier03},
 the CFRS survey \citep{lilly95II}, and
 from unpublished analysis of CFHT emission line searches in the CFRS
 22h field \citep[][]{cramlilly99,tran04}.
The \OII\, lines of the CADIS galaxies were measured using a
Fabry-Perot interferometer at about 9200\am\, \citep[for details,
  see][]{hippe03,maier03}.  The CFHT  survey was undertaken combining multiple parallel long slits with a narrowband
filter to search for $z\sim 6.6$ Lyman $\alpha$ emitters in the 9200 {\AA}
atmospheric window \citep[for details see][]{cramlilly99,tran04}. These CFHT \OII\,-selected
galaxies are located in the CFRS 22h field \citep{lilly95II}.

~~~ The total sample of $z\sim 1.4$ galaxies contains five objects,
 two from the CADIS, one from the CFRS survey, and two from the CFHT
 multislit$+$narrowband filter survey.
 For all
 galaxies we obtained VLT-ISAAC spectra in the J band, to measure the
 \Hb\, and \OIIIa\, fluxes, and in the H-band, to measure \Ha\, and
 (upper limits to the) [NII] fluxes. 
  We also acquired H- and J-band
 imaging in order to test the relative calibration of the emission
 line fluxes, as described in Paper\,III.
Fragmentary near-infrared spectroscopy data were obtained for three
additional objects, and we will report these elsewhere if we
made up the remaining flux measurements.

\begin{figure}[h!]
\plotone{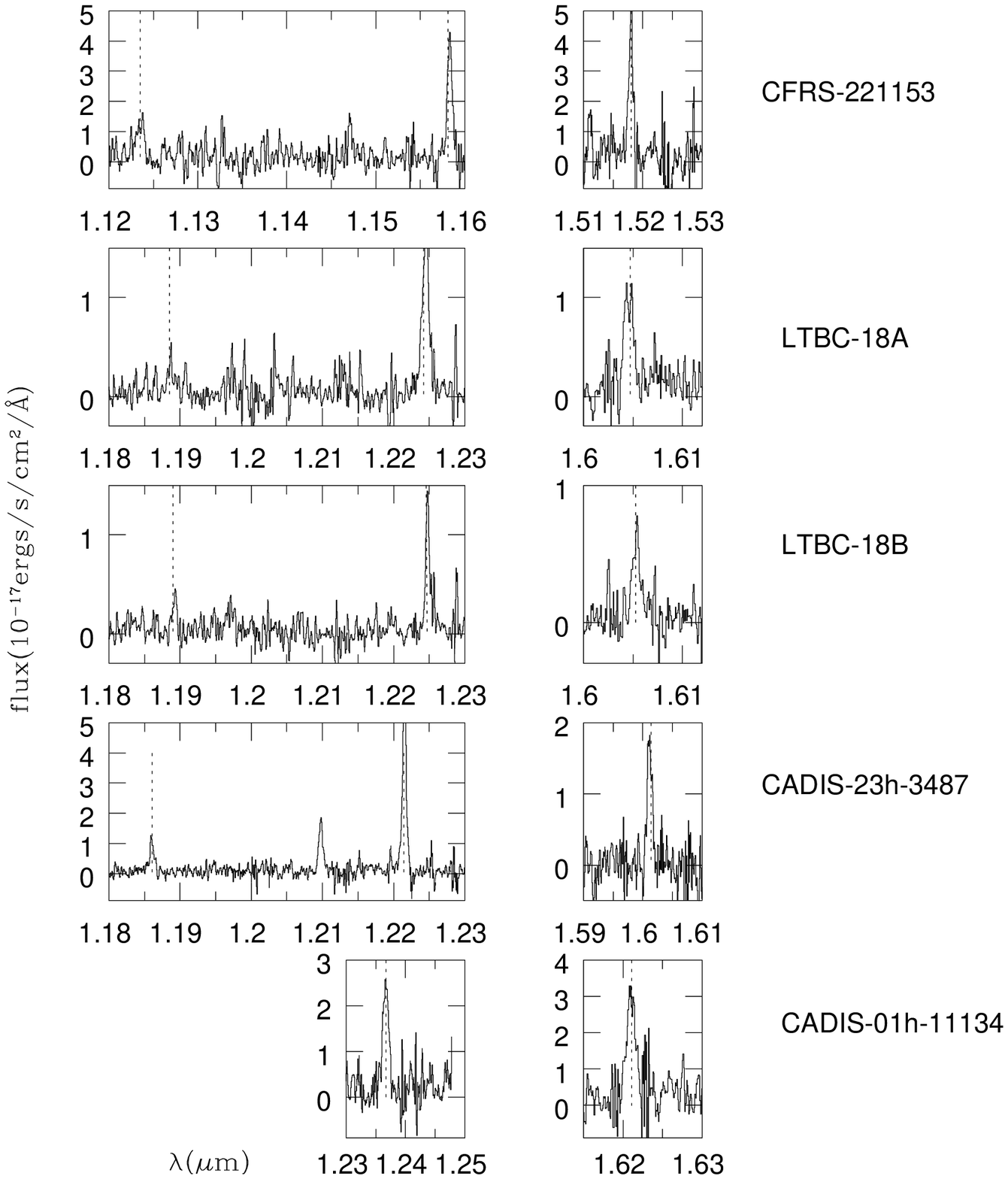}
\caption
{\label{Spectra1D} \footnotesize The near-infrared spectra of the five $z\sim1.4$
  galaxies showing  \Hb\, and \OIIIa\, from J-band spectra (left panel), and \Ha\,
  from H-band spectra (right panel). The location of the respective
  emission line is shown by  vertical dashed lines.
Note that the  \Hb\, line  of CADIS-01h-11134 coincides with a strong OH sky line,
so the respective flux could not be measured. 
}
\end{figure}

~~~ The observations were carried out at the VLT in visitor mode in October 2004
 (Program 074.B-0122A) using a slit of 2\arcsec\, width and
 120\arcsec\, length. The spectra cover a range of 79\,nm in the H band, and 59\,nm in the J
 band.  The individual integration times and the
 standard stars used for the flux calibration are given in Table
 \ref{ObsISAACz14}. The data reduction was done using IRAF as described in detail in
 Paper\,III for similar ISAAC spectroscopic observations of CFRS
 galaxies at medium redshift. The resulting spectra of the five
 $z\sim1.4$ galaxies are shown in Figure\,\ref{Spectra1D}. 
 For easy reference we use throughout this paper
 the galaxy identification codes adopted in the previous quoted
 literature.
 
~~~ In the good seeing conditions for the ISAAC observations (see Table~\ref{ObsISAACz14}),
the 2\arcsec\, ISAAC slit approximates a total flux measurement. It is
well matched to the 1\arcsec.75  width of the CFRS observations for CFRS-221153 \citep{lef95} and to the 2\arcsec\,  width of the slit used
for the observations of LTBC-18A and LTBC-18B \citep[][]{cramlilly99,tran04}. The \OII\, measurements for the two CADIS objects were made with  an imaging  Fabry-Perot interferometer \citep[see, e.g.,][]{maier03} and thus the measured flux  should be ''total".  In Paper\,III
we were able to check directly the relative flux calibration of optical
and near-infrared spectra, under similar observational set-ups, and
concluded that there was no evidence of problems. One possible concern
with the present data is that the slits used by \citet{cramlilly99} may
not have been accurately centered on the two LTBC objects, conceivably
leading to an underestimate of the \OII\, flux for these particular
objects.  However, these two objects have similar emission line ratios to the other $z \sim 1.4$ galaxies and slit mis-centering is not believed to be a problem.

\subsection{Comparison samples}
\label{compsampl}

In this paper we compare our new $z\sim1.4$ metallicities data with
the properties of 30 CFRS galaxies at  $0.47<z<0.92$  from Paper\,III,
and  with  17 lower luminous $0.4<z<0.7$ CADIS galaxies from
\citet{maier04}.
Although only few of the $0.4<z<0.7$ CADIS galaxies have five emission
lines observed, \citet{maier04} showed that it is likely that all 17   CADIS
galaxies have low extinction, so we  used the case B intensity ratio
of \Ha\, to \Hb\, of 2.86 when running our program to determine
metallicities based on the KD02 models. 
As nearby comparison samples we adopted, as in Paper\,III,  70 $z<0.095$ KISS
galaxies of \citet{melbsal}, and  108 NFGS $z<0.04$ galaxies of
\citet{jansen}, with  all \OII, \Hb, \OIIIa, \Ha, and [NII] line fluxes
measured.

~~~ After careful analysis, we opted for \emph{not} using, in our analysis of the
metallicities, the following samples of galaxies with published oxygen
metallicities: the 64 $0.26<z<0.82$ galaxies of \citet{kob03}, the 177
$0.3<z<1$ galaxies of \citet{kobkew04}, the 25 $0.4<z<0.9$ galaxies of
\citet{liang04}, and the 7 galaxies at $2.1<z<2.5$  of \citet{shapley04}.
In fact, from all these samples, only a handful of low redshift
(typically $z\lta0.5$) objects had \Ha\, and \NII\, measurements, which are important
to break the \R23\, degeneracy with metallicity. 
In detail, 9/66 galaxies at $0.26<z<0.82$ with measured
  equivalent widths  from \citet{kob03}  have measured
  \NII/\Ha\, (all at $z<0.4$);
32/177 galaxies at $0.3<z<1$ with measured equivalent widths
  from \citet{kobkew04}  have measured \NII/\Ha\, (all at $z<0.5$);
and 4/25 galaxies at $0.4<z<0.9$ from \citet{liang04}   have \NII/\Ha\, measured.
Furthermore, several of the
mentioned studies performed no dust extinction correction. This, as we showed
in Paper\,III, can significantly affect the analysis. Finally, the $2.1<z<2.5$
galaxies of \citet{shapley04} had oxygen abundances determined using only the \NII/\Ha\, ratio;
these, as already discussed by \citet{shapley04} and in Paper\,III, are on their own rather
uncertain \citep[see also][]{dokkum05}.
We do not think that excluding these samples from our discussion would
significantly change our conclusions. Moreover, as discussed below, using the CADIS and
CFRS comparison samples has several advantages.

\subsubsection{Advantages of the CADIS and CFRS comparison samples}

The choice of the CADIS and CFRS samples for comparison with our new
set of metallicities for galaxies at $z\sim1.4$ has  two main advantages:

~~~ i) As stressed by \citet{salzer05}, the details of any
specific calibration of line fluxes and ratios into metallicity has a
significant impact on the metallicity measurements and the resulting luminosity-metallicity relation. It is
thus important that, when comparing different samples of galaxies, identical
methods and calibrations are adopted for all samples \citep[see also][]{elkew}.  For all the galaxies of
the adopted comparison samples we had access to the emission line
fluxes measurements and derived [O/H] metallicities using the same
approach that we used for calculating the metallicities of the $z\sim1.4$
galaxies (see Section \ref{OHSFRsAV}, and Paper\,III for details on the method).

~~~ ii) Another important requirement for performing a proper comparison of the
metallicity-luminosity relation between low- and high-z samples of galaxies is
the availability of properly derived rest-frame luminosities for the distant
galaxies. As also stressed by \citet{salzer05}, one should fit the full SEDs
over a large wavelength range in order to obtain reliable rest-frame
luminosities for the $z>0$ galaxies. We have available multi-wavelength
imaging for the $z\sim1.4$ galaxies presented in this paper, as well as for
the CFRS galaxies of Paper\,III (see Table\,1 of Paper\,III), and the CADIS
objects where 16 filters were used to derive the SED \citep[see][]{hippe03,maier04}.  Reliable rest-frame luminosities could
therefore be obtained for all the galaxies in these samples.

\subsubsection{Selection effects  in the CADIS and CFRS  samples}
\label{selef}

~~~ The CADIS and CFRS galaxies at $0.4<z<1$ have
different selection criteria, but, excluding a few objects,  we can get a useful comparison sample,
as shown in the following. 
The CFRS galaxies are selected to have $I_{AB}<22.5$,
and  a rest frame  equivalent width of \Hb, $EW_{0}(\Hb)$, larger than
8\,\AA, as described in Paper\,II. 
On the other hand, the CADIS galaxies were selected based on
their \OIIIa\, emission detected using a Fabry-Perot-Interferometer ($\rm{F}_{\rm{lim}} = 3 \cdot 10 ^ {-20}
 \rm{W}  \rm{m}^{-2}$), regardless of their continuum
brightness.
Thus, CADIS can detect lower luminosity  galaxies at a given $z$.

\begin{figure}[h!]
\plotone{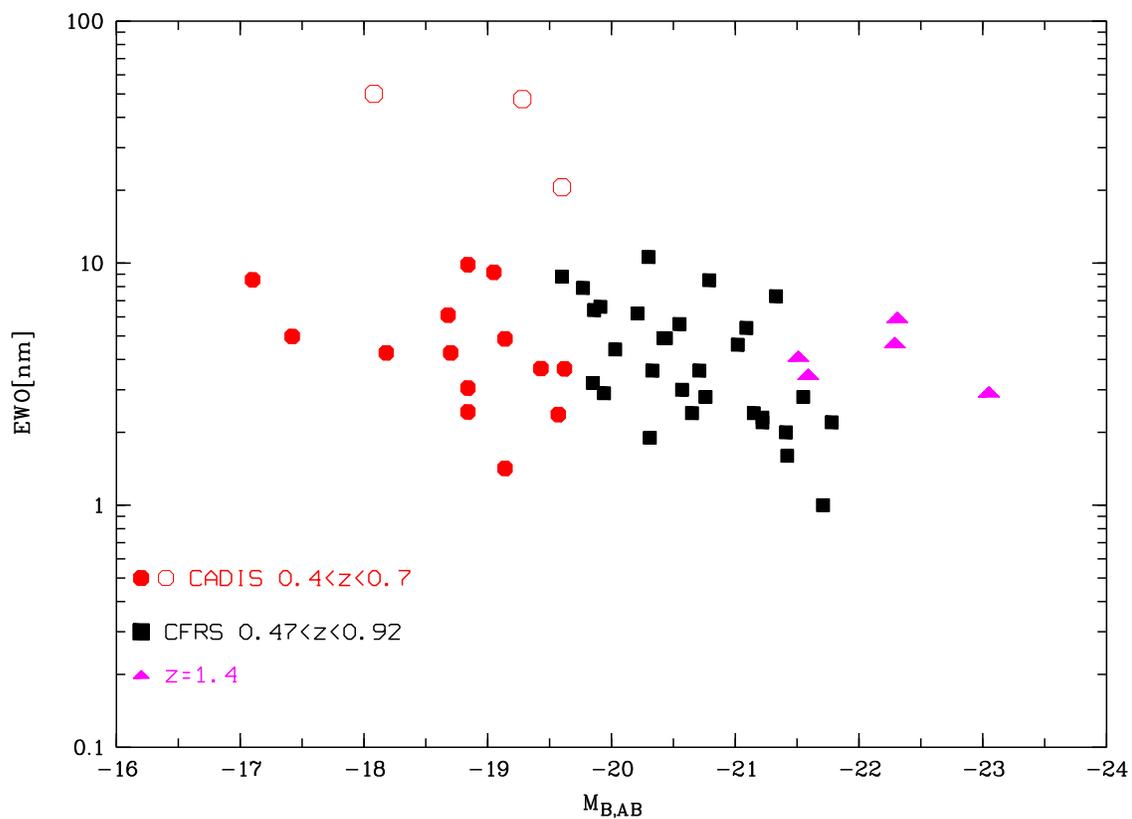}
\caption
{\label{EWMB} \footnotesize Line equivalent widths (EWs) at rest wavelength
  versus absolute blue magnitude for the $0.4<z<0.7$ CADIS galaxies and
  $0.47<z<0.92$ CFRS galaxies in our comparison samples, and for the
  $z\sim1.4$ galaxies.
The distribution of the EWs of CADIS objects is consistent with the EW
  distribution of the CFRS sample and $z\sim1.4$ galaxies, except three CADIS  objects at
  $z\sim 0.4$ with particular high equivalent widths (open circles), which are excluded from the comparison sample for the further discussion. 
}
\end{figure}

Fig.\,\ref{EWMB} shows the distribution of  EWs at
restframe wavelength of $z\sim1.4$, CADIS and CFRS galaxies as a function of blue
absolute magnitude. The distribution of EWs of the CADIS galaxies is
broadly consistent with the EW distribution of CFRS and $z\sim1.4$ galaxies \citep[and this
holds  also for the larger CADIS sample as shown in Fig.\,4
in][]{hippe03}, except for three
CADIS objects at $z\sim0.4$ (open circles) which have a particular high EW. These are
the only galaxies without VLT spectroscopic follow-up reported by \citet{maier04}, 
galaxies  with  high enough fluxes that their  oxygen abundance could be obtained using
spectroscopy with the  3.6-m  Telescopio Nazionale Galileo (TNG). 
We will exclude these three objects from
the CADIS comparison sample in the following.

~~~ Summarizing the above, despite the different
selection criteria,  the $M_{B,AB}<-21.5$ $z\sim1.4$ galaxies, the $M_{B,AB}<-19.5$ CFRS, and the 
 $-17.5>M_{B,AB}>-19.5$ CADIS sample follow a similar EW-$M_{B,AB}$ distribution
 (after excluding the three objects mentioned above) suggesting that they are not completely
different species and that the CADIS+CFRS sample is a useful comparison
sample for the $z \sim 1.4$ galaxies.  It is not perfect, but it is the
best we can do at the present time. 
The CADIS objects were [OIII] selected, whereas the CFRS is effectively
\Hb\, selected (see Paper\,II). It should be appreciated that CADIS
will be slightly biased towards high [OIII]/[OII] (and low [O/H]),
inferred from the fact that there is a correlation within the CFRS sample between
$L_{[OIII]}$ and [OIII]/[OII], and $L_{[OIII]}$ and [O/H].
There is no such correlation with $L_{[OII]}$, indicating that there
should not be similar biases in the $z\sim1.4$ \OII\, selected sample.

%
%

\section{The measurements}

\subsection{Emission line fluxes}

~~~ Emission line fluxes were measured from the calibrated spectra
following the procedure described in Paper\,III. Table
\ref{VLTISAACz14} reports, for each individual galaxy, the emission
line fluxes (or upper limits) of the five observed galaxies.
Specifically:

~~~ {\it (i)} The \NII\, line was too faint to be detected. However, we
could determine upper limits to the \NII\, flux for all five $z\sim1.4$ galaxies, which are good enough to check for
contamination by AGNs and to break the degeneracy of the \R23\, relation.
 Table \ref{VLTISAACz14} lists the
$2\sigma$ upper limits.

~~~ {\it (ii)} For  four of the $z\sim1.4$ galaxies we were able to
measure fluxes for all three lines of \Hb, \OIIIa, and \Ha.
The fluxes and flux errors were measured as described in
  Paper\,III. We have not applied any
  corection for stellar absorption to the \Hb\, line, because this line  has an observed equivalent width
larger than 50\AA\, for the four $z\sim1.4$ galaxies. Like in Paper\,III, we assumed a minimum 10\% for
  the uncertainties of the measured fluxes when computing oxygen
  abundances using the KD02 models (see Table\,\ref{VLTISAACz14}).

~~~ {\it (iii)} For the galaxy CADIS-01h-11134 the \Hb\, line coincides with a strong OH
line, so we were not able to  measure the \Hb\, line flux.
However, we could measure \OII, \OIIIa, \Ha, and an upper limit for
\NII\, for this galaxy. We calculated the \Hb\, flux for the galaxy  CADIS-01h11134 using case B recombination and assuming a low
$A_{V}=0.5$  (similar to the extinction values  found for the other
four $z\sim1.4$ galaxies), and we used
this value when calculating the oxygen abundance of this galaxy using
the models of KD02 (see Section\,\ref{OHSFRsAV}).
This  results in a value of
$A_{V}=0.39^{+0.43}_{-0.39}$, which is completely consistent with the input
value. This makes the point that $A_{V}$ is not solely determined by
\Ha/\Hb, but also has a small contribution from the other line ratios.


\subsection{Gas metallicities, SFRs and extinction parameters}
\label{OHSFRsAV}

~~~ The approach described in Paper\,III was used to derive [O/H] and the
extinction values $A_V$ for the five  $z\sim1.4$ galaxies.  Briefly,
the approach is based on the models of KD02, who developed a set
of ionisation parameter and oxygen abundance diagnostics based on the use of
strong rest-frame optical emission lines.  The method consists in performing a
simultaneous fit to all available emission line fluxes (including the
\NII\, upper limits) in terms of extinction
parameter $A_{V}$, ionisation parameter $q$, and [O/H].

\begin{figure}[h!]
\plotone{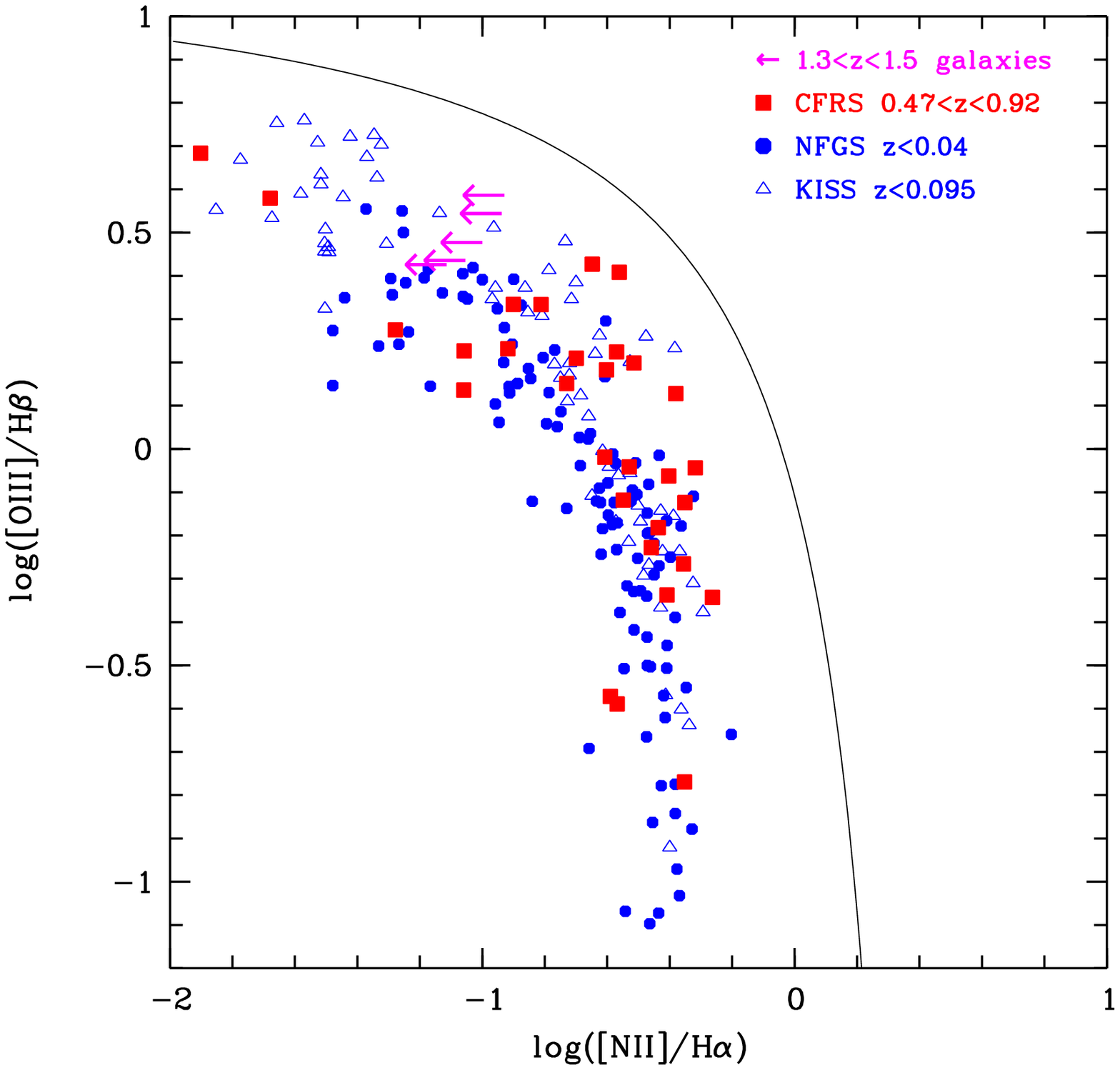}
\caption
{\label{diagnAGN} \footnotesize Diagnostic diagram to disentangle
  star-formation dominated galaxies from AGNs. The position of the
  five galaxies at $z \sim 1.4$ is shown by the arrows, since only
  upper limits for \NII\, could be derived. Also plotted are the 30
  $0.47<z<0.92$ CFRS galaxies of Paper\,III (filled squares), 108 local
  NFGS galaxies from \citet[][filled circles]{jansen}, and 70
  local KISS galaxies from \citet[][open triangles]{melbsal}.  All the
  plotted galaxies, including the $z\sim 1.4$ galaxies, lie below the
  solid line, i.e., the theoretical threshold computed by
  \citet{kewley01} above and to the right of which galaxies are
  dominated by an AGN. In all plotted galaxies the emission is thus
  dominated by star formation.  }
\end{figure}

~~~ As in Paper\,III, before proceeding with determining gas
metallicities, we checked that the detected line emission is not
strongly contaminated by the presence of an AGN.  To this purpose we
used the log(\OIIIa/\Hb) vs.  log(\NII/\Ha) diagnostic diagram shown
in Fig.\,\ref{diagnAGN}.  The location on this diagram of the five $z\sim1.4$
galaxies is indicated with arrows, given that we only had
upper limits for the [NII] fluxes.  The five galaxies lie under and to
the left of the theoretical curve of \citet[][solid line]{kewley01},
which separates star forming galaxies (below/left of the curve) from
AGNs (above/right of the curve). In none of the five $z\sim1.4$ galaxies
the line emission is therefore dominated by an AGN.

~~~ The gas metallicities we then derived for the 
  $z\sim1.4$ galaxies are rather low compared to lower redshift
  galaxies of similar luminosities,
and also their derived  extinction values  are 
quite low. We will discuss this finding in the following
Sect.\,\ref{discuss}.
 The \Ha\, luminosities  (corrected for
extinction)   were used to calculate
the SFRs of the five $z\sim1.4$ galaxies using the \citet{ken98} conversion of H$\alpha$ luminosity into
$\dot{\rm{M}}$: $\rm{SFR} (M_{\odot}\rm{yr}^{-1}) = 7.9 \times 10^{-42}
\rm{L}(\rm{H}\alpha)\rm{ergs/s}$. The resulting SFRs are given in  Table \ref{VLTISAACz14}.
 Note that  alternative (but less
probable)  oxygen abundance solutions are found for two $z\sim1.4$ galaxies, and we will indicate
them as open squares wherever [O/H] is plotted.
These alternative (but less probable) oxygen abundances found using the method described in
Section 3.3  of Paper\,III are [O/H]=$8.02^{+0.10}_{-0.07}$
for object LTBC-18A, and [O/H]=$8.34^{+0.16}_{-0.12}$ for object  CADIS-23h-3487.

%
%

\section{Discussion}
\label{discuss}

\subsection{Basic observational result}
\label{basicobs}

~~~ Our basic observational result is that the spectra of the $z \sim 1.4$
galaxies, which have $-21.5 \ga M_{B,AB} \ga -23$ have the line ratios
and derived parameters (metallicity, extinction, ionisation parameter)
that are found in much lower luminosity galaxies at lower redshifts. Specifically,
they are similar to what is found in the $0.4<z<0.7$ CADIS
galaxies ($M_{B,AB} > -19.5$), and noticably different from the general properties of the more luminous
($-19.5 > M_{B,AB} > -22$) CFRS
galaxies at $0.47<z<0.92$ (see Paper\,III).

\begin{figure}[h!]
\plotone{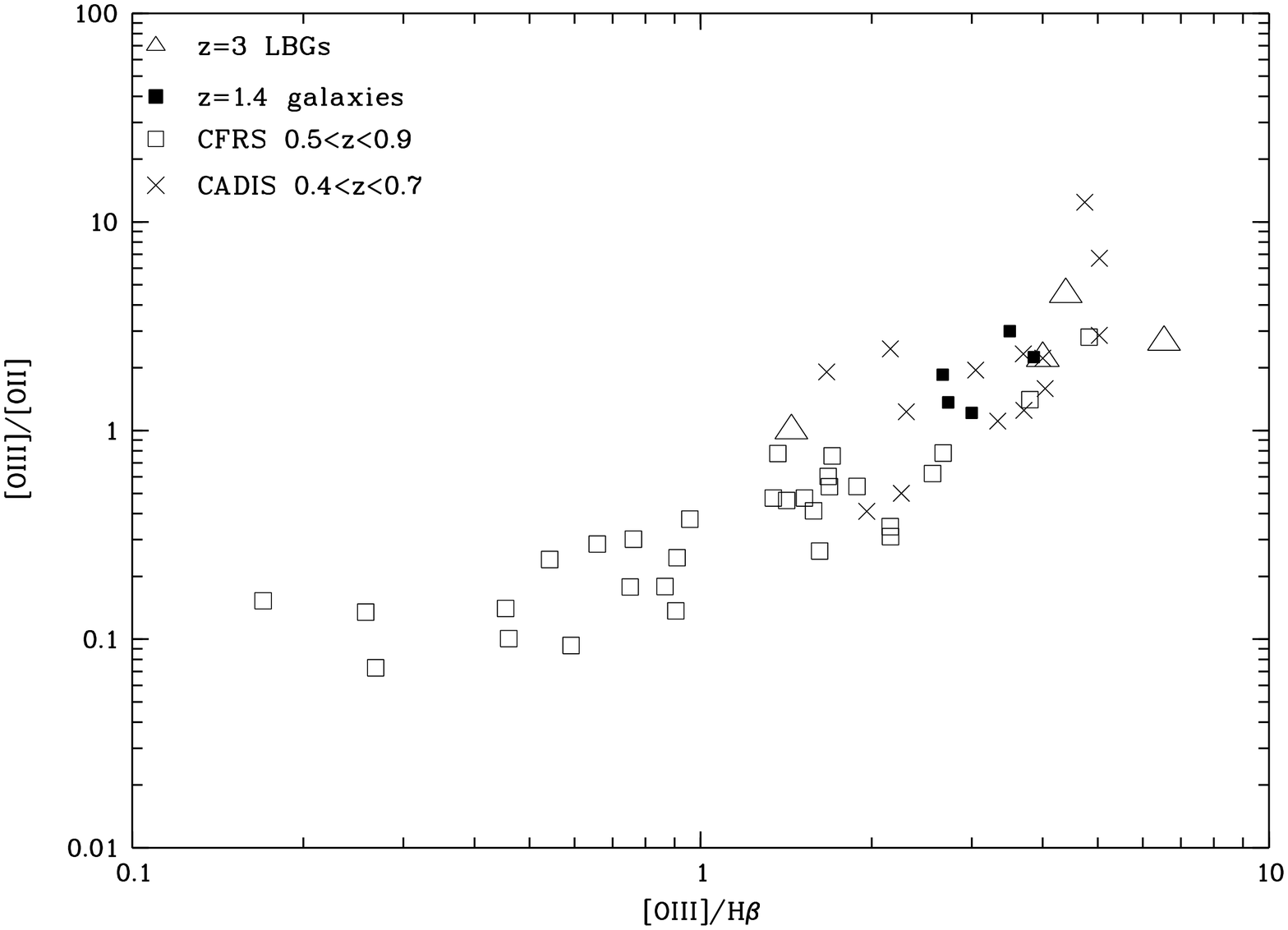}
\caption
{\label{diagnOIIIOII} \footnotesize  Line ratios \OIIIa/\OII\, vs. \OIIIa/\Hb\,
  for the  $z\sim 1.4$ galaxies (solid squares). Also plotted are
  the 30 $0.47<z<0.92$ bright CFRS galaxies (open squares), the $0.4<z<0.7$ CADIS galaxies
  of \citet[][crosses]{maier04}, and the $z\sim3$ LBGs galaxies
  (open triangles). The five $z\sim 1.4$ galaxies have the high \OIIIa/\OII\,
  ratios that are typical of the lower luminosity CADIS galaxies and
  of the more luminous higher-$z$ LBGs.
}
\end{figure}

We look first at the simple  line ratios.  
Fig. \ref{diagnOIIIOII} shows the location of the five $z\sim
1.4$ galaxies on the diagnostic diagram \OIIIa/\OII\, (extinction
corrected) vs. \OIIIa/\Hb.  The
lower luminosity $0.4<z<0.7$ CADIS galaxies, the brighter $0.47<z<0.92$ CFRS galaxies, and
the more luminous $z\sim 3$ LBGs are also plotted for comparison.  In contrast with the
bright, intermediate-z CFRS galaxies, which show low \OIIIa/\OII\, and
\OIIIa/\Hb\, ratios, it should be noted that all the $z\sim1.4$
galaxies have the high \OIIIa/\OII\, and \OIIIa/\Hb\, line ratios which
are typical of the lower luminosity intermediate-$z$ CADIS 
galaxies and of the more luminous LBGs at higher $z$.
In other words, the appearance of high  \OIIIa/\OII\, ratios moves to
brighter luminosities at higher $z$.

\begin{figure}[h!]
\plotone{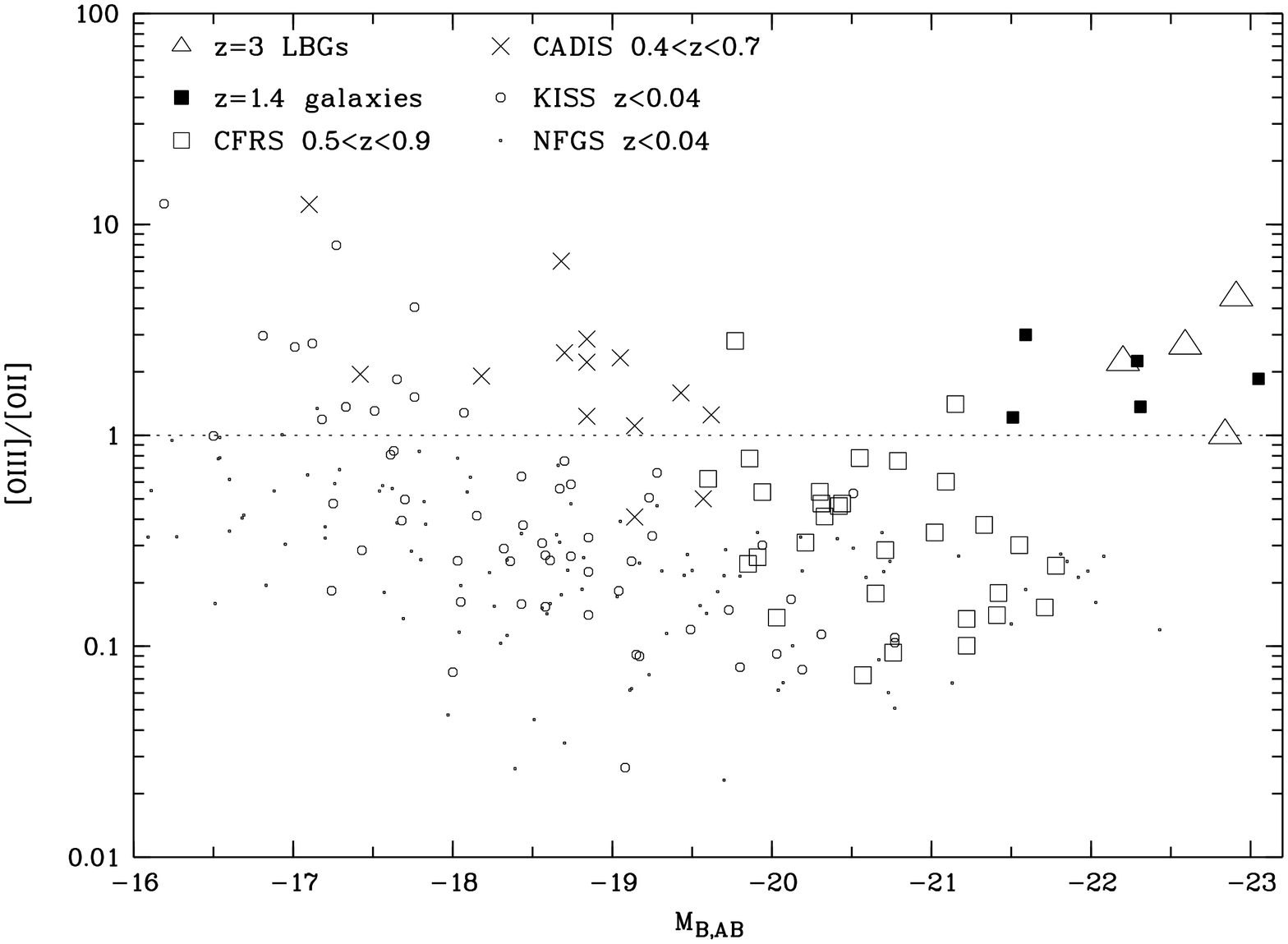}
\caption
{\label{OIIIOIIMB} \footnotesize  \OIIIa/\OII\, vs. $M_{B,AB}$
  for galaxies at $0<z<1.5$ with symbols as in
  Fig.\,\ref{diagnOIIIOII}, with additional plotted  local KISS (open circles) and NFGS (dots) galaxies.
The basic trends in our data are consistent
with a ``downsizing'' picture \citep{cowie96} in the sense that a given type 
of spectrum (high  \OIIIa/\OII) are found almost exclusively below a luminosity threshold
which is low at zero redshift ($M_{B,AB} \sim -18.5$), increases to 
$M_{B,AB} \sim -20.5$ at $z \sim 0.7$, and has evidently increased to
above $M_{B,AB} \sim -23$ at $z \sim 1.4$.}
\end{figure}

To illustrate this further, Fig. \ref{OIIIOIIMB} shows the line 
ratios \OIIIa/\OII\, vs. $M_{B,AB}$
  for the five $z\sim 1.4$ galaxies, the CADIS and CFRS sample,  
  the LBGs, and the local KISS and NFGS galaxies. 
It is clear that there is a relation between \OIIIa/\OII\, and  $M_{B,AB}$ for
local KISS galaxies, in the sense that lower luminosity galaxies tend
to have higher \OIIIa/\-\OII\, ratios.  At higher redshifts, a similar
trend is seen for the combined sample of CADIS and  CFRS
galaxies at $0.4 \la z \la 0.9$, with the relation displaced upwards and to the
right, i.e. 
to higher \OIIIa/\OII\, ratios at a given luminosity or higher luminosities
at a given line ratio.  
The $z\sim1.4$ and LBG galaxies at $z\sim3$ continue this
trend, uniformly exhibiting the
high \OIIIa/\OII\, ratios (i.e. above one). While the \OIIIa/\OII\,
distribution of the \OIIIa\, selected CADIS objects may be biased (see
Section \ref{selef}),  the $z \sim 1.4$ galaxies
were \OII-selected and there is no observational selection against low \OIIIa/\OII\,
ratios.

The tracks of individual galaxies in this diagram are likely to be diagonal,
moving down to lower \OIIIa/\OII\, with modest fading.  Within the population, the
evolutionary trend is in the sense that, as we look back
in time, galaxies of similar luminosities as local galaxies show
increasingly higher \OIIIa/\OII\, ratios.
We do not know at this point about the spectra of low luminosity galaxies at 
high redshifts.
It can be nevertheless seen that a  given type of spectrum (high 
\OIIIa/\OII\,) is found almost exclusively below a luminosity threshold
which is low at zero redshift ($M_{B,AB} \sim -18.5$), increases to 
$M_{B,AB} \sim -20.5$ at $z \sim 0.7$, and has evidently increased to above
$M_{B,AB} \sim -23$ at $z \sim 1.4$.
One may   interpret these basic trends seen in our data within a
"downsizing" picture" \citep{cowie96}.

''Down-sizing'' as originally introduced by \citet{cowie96} defined
"forming galaxies" as those in which the timescale of formation M$_{*}$/SFR
was less than the Hubble time at the redshift in question. \citet{cowie96} then
showed that there existed a threshold in mass below which forming
galaxies were found and above which they were not found.  "Down-sizing"
then referred to the observational fact that this mass threshold
decreases (in mass) with cosmic time.  This concept has since been
generalised to any situation in which signatures of "youthfulness" (e.g.
irregular morphologies, blue colours, high equivalent width emission lines, and, as  earlier argued in Paper\,II, low [O/H]) exhibit a similar mass (or more loosely luminosity) threshold which decreases with epoch.

\begin{figure}[h!]
\plotone{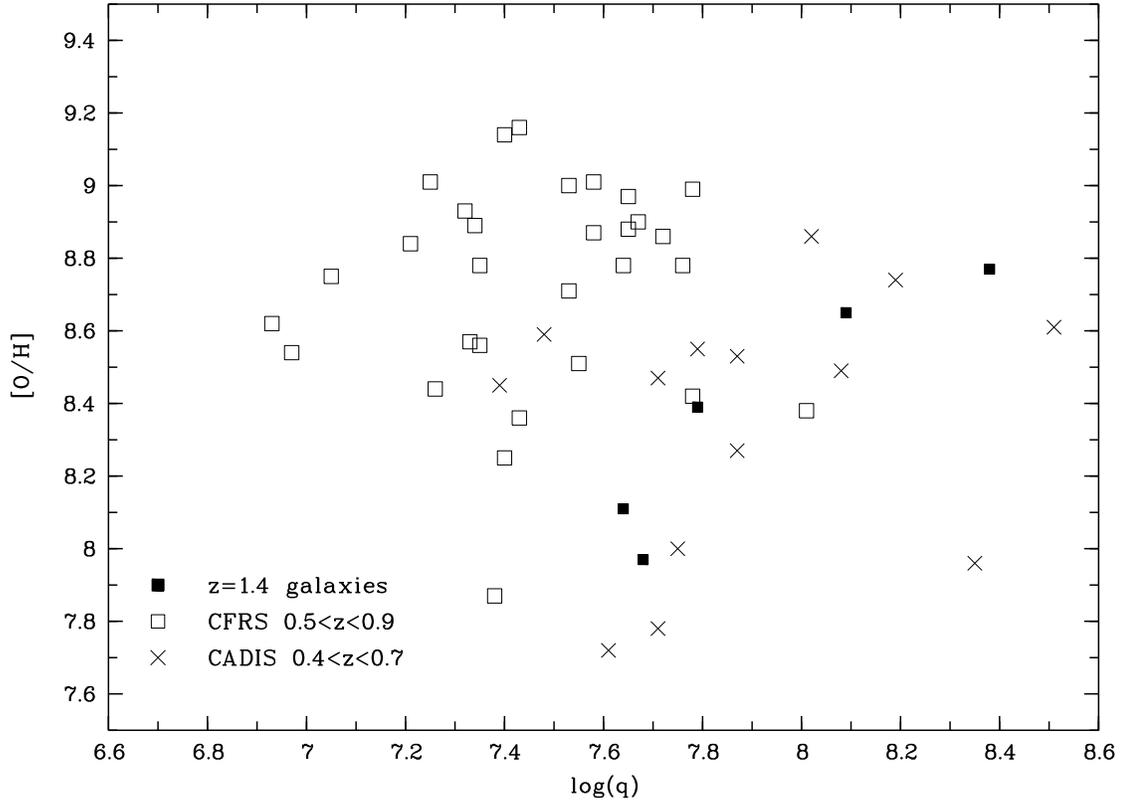}
\caption
{\label{OH_logq} \footnotesize  Oxygen abundance vs. ionisation  parameter  
for galaxies at $0<z<1.5$ with symbols as in Fig.\,\ref{diagnOIIIOII}.
The $z \sim 1.4$
galaxies overlap in $q$ and [O/H]
the CADIS objects, and avoid the area defined by the more luminous CFRS
galaxies with lower $q$ and higher [O/H], showing that the high
\OIIIa/\OII\, ratios shown in Fig.\,\ref{OIIIOIIMB} arise from either or both  high ionisation
parameter $q$ and low metallicities.
}
\end{figure}

High \OIIIa/\OII\, ratios arise from a  high ionisation
parameter $q$ and/or low metallicities, as shown, e.g., in KD02, their
Fig.\,1.
This is also illustrated in Fig.\,\ref{OH_logq},
which shows the oxygen abundance as a function of ionisation parameter
for the $0.4<z<1.5$ galaxies.  It can be seen that the $z \sim 1.4$
galaxies overlap in $q$ and [O/H]
the CADIS objects, and avoid the area defined by the more luminous CFRS
galaxies with lower $q$ and higher [O/H].
In the next sections, we explore more directly the implied metallicities
of the galaxies in our different samples.

%
%

\subsection{The P\'egase2 chemical evolution models}

~~~ To quantify our results for the chemical evolution of
galaxies with cosmic time, we have constructed a grid of chemical evolution
models using P\'egase2 \citep{fiocrocca99}.
P\'egase2 is a user-friendly galaxy evolution code that computes
galaxy properties as a function of galaxy age, starting from the
properties of simple stellar populations (SSPs), i.e. populations of
stars formed simultaneously with the same metallicity. Metallicities
between 0.005 and 5 times solar are implemented in the code.  The
input parameters for P\'egase2 are:

\noindent
{\it (i)} $\varphi(M)$, i.e., the shape of the stellar initial mass
function (IMF).  We adopted a Salpeter IMF, with $\alpha=-2.35$
between 0.1 and 120 $M_{\odot}$.

\noindent
{\it (ii)} $Y$, i.e., the chemical yields from nucleosynthesis.    
We used \citet{woosw} B-series models for massive
stars.

\noindent
{\it (iii)} $M_{tot}$, i.e., the total mass of gas 
 available to form
the galaxy.

\noindent
{\it (iv)} $t_{infall}$, i.e., the timescale on which the galaxy is
assembled.  The model assumes that galaxies are built by continuous infall of
primordial (zero metallicity) gas with an infall rate that declines
exponentially as: 

\begin{equation}
\dot{M}(t) = (M_{tot} \cdot e^{-t/t_{infall}}) / t_{infall}. 
\end{equation}

\noindent  The total mass of the galaxy at a given time $t$ is then
given by:

\begin{equation}
M(t) = M_{tot} \cdot (1- e^{-t/t_{infall}}).
\end{equation}

\noindent
{\it (v)} $\psi(SFR)$, i.e., the functional form of the star formation
rate.  We adopted a star formation rate of the form:

\begin{equation}
\psi(t)  = M_{gas}(t) \cdot p_2 \cdot e^{-t/t_1} / t_1,
\end{equation}

\noindent 
where $p_2 \cdot e^{-t/t_1} / t_1$ respresents a kind of star
  formation ``efficiency'', in the sense that it is $SFR(t) /  M_{gas}(t)$.
  The  mass of the gas in the model galaxy at a time $t$,
$M_{gas}(t)$, is equal $M(t)-M_{*}(t)$, where $M_{*}(t)$ is the mass locked up into
stars at the time $t$. 
Models with $t_{infall}>t_1$ are not very likely, since the
star formation effectively ceases, even though gas is still infalling. We therefore
restricted our grid to models with $t_{infall} \le t_1$.

\noindent
{\it (vi)} $A_V$, i.e., the extinction due to dust.  The
``inclination-averaged'' extinction prescription that is implemented in
P\'egase2 was included in the construction of our models.

~~~ To summarize, the set of  P\'egase2 models we are using consists of the following: it is
assumed that there is a reservoir of primordial  gas  that feeds the
``galaxy''.
The galaxy forms stars with a rate $\psi(t) = (M_{gas}(t) \cdot
p_2 \cdot e^{-t/t_1}) / t_1 $, and we follow the evolution of the
model galaxy in the luminosity-metallicity diagram as a function of time.
 The model galaxy is treated as a single zone with uniform
chemical composition and without outflow. This oversimplification is appropriate to get a
basic understanding of how fundamental parameters, such as SFR and gas
supply, roughly affect galaxy properties such as luminosity and metallicity.

We stress that we do not assign any particular physical significance to these models,
but simply use them as means to explore reasonable possibilities. By
varying the parameters we can get an understanding of how different
evolutionary histories are represented in the [O/H]-$M_{B}$ plane.
%

%


\subsection{Constraints on the chemical evolution of galaxies}
\label{constrmodels}

\begin{figure}[h!]
\plotone{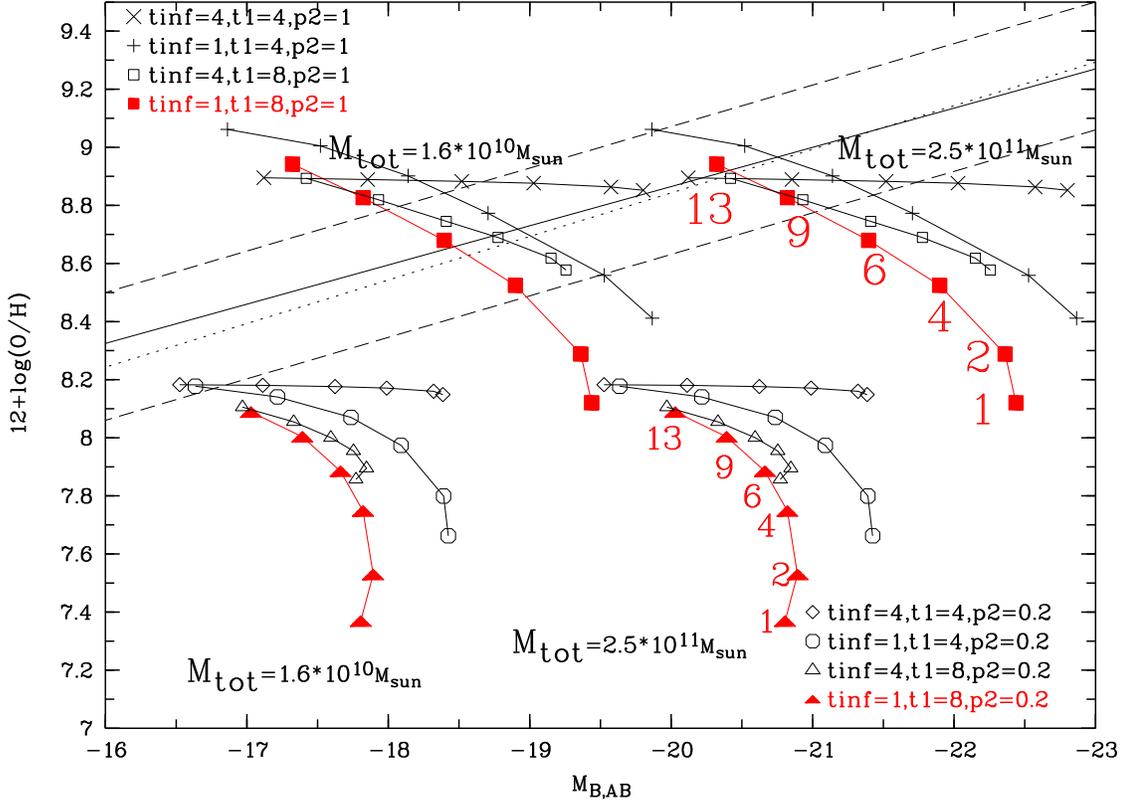}
\caption
{\label{MB_OHPegas} \footnotesize Several P\'egase2 models for different
  $t_1$, $p_2$, $t_{infall}$, and $M_{tot}$, compared to the local
  metallicity-luminosity relation. The
  nearby NFGS and KISS galaxies have oxygen abundances lying between the dashed lines
  (with the solid and dotted lines indicating the mean local relation
  for NFGS and KISS galaxies, respectively). }
\end{figure}

~~~ We constructed a large grid of P\'egase2 models to explore which region of the
parameter space could reproduce the constraints imposed by the local
metallicity-luminosity relation and by the metallicities and luminosities of galaxies at
higher redshifts. To build the grid, we varied $t_1$ (in Gyrs),
$p_2$ (dimensionless), $t_{infall}$ (in Gyrs), and $M_{tot}$ (in solar
masses).  
Fig.\,\ref{MB_OHPegas} shows the relation between the gas abundance
12+log(O/H) versus the absolute B magnitude (in the AB system) for a
set of 
different illustrative models in the grid.  In the figure, the symbols along each track
indicate the age of the model galaxy, varying from 1 to 13 Gyrs from 
bottom to top of each track. With reference to the figure the following
should be noted:

\noindent 
{\it (i)} Models with  short $t_{infall}$ and $t_{infall}<<t_1$ in Fig.\,\ref{MB_OHPegas} (e.g.,
$t_{infall}=1$ and $t_1=8$)
are models in which the mass of gas in the reservoir of gas is
transferred to the galaxy   in a short time,  after which the
gas infall practically stops, resulting in gas rich galaxies at early times.
The galaxy subsequently forms stars in a roughly ``closed-box''-like scenario.

\noindent 
{\it (ii)} Increasing  $t_{infall}$ (but still keeping $t_{infall}<t_1$,
e.g., $t_{infall}=4$ and $t_1=8$) results in models in which  the
galaxy reaches a high metallicity  at early times. 
Ultimately, models with $t_1 = t_{infall}$ reach
their end metallicity after $\sim1$\,Gyr, and  afterwards their metallicity
remains approximately constant while the luminosity decreases. This results in
an almost horizontal evolution in the luminosity-metallicity diagram
with [O/H] effectively constant over a long period of time.

\noindent 
{\it (iii)} Varying $M_{tot}$ results in an horizontal shift of the models
as shown in Fig.\,\ref{MB_OHPegas} for two typical masses 
($M_{tot}=2.5\times 10^{11}M_{\odot}$, and $M_{tot}=1.6\times 10^{10}M_{\odot}$).

~~~ Searching for  P\'egase2  models    consistent with the observed
metallicities and luminosities of galaxies, we notice 
a constraint for the  models given by the observed properties of
local galaxies:   the evolution
of the  model  galaxies in the metallicity-luminosity diagram, following
the  P\'egase2 tracks, has to lead  to the region occupied by  local galaxies today.
If galaxies started at very high $z$ they get in the region of the
metallicity-luminosity relation today  after  about 13\,Gyrs. But they could get there at younger ages if they started
their evolution  well after the Big Bang (see below).

The existence of relatively high luminosity galaxies with relatively
low metallicities implies that the horizontal ``constant-metallicities''  models
described in (ii) are unlikely.  These  galaxies would have to fade  too
much  following the horizontal models (ii) to reach the local
metallicity-luminosity relation: e.g.,  in Paper\,II in Sections 4.1 and 4.2  it
was discussed that it is unlikely that CFRS galaxies can fade by more
than 2.5\,mag to the present epoch.

Thus, a key point seen in  Fig.\,\ref{MB_OHPegas} is that galaxies which
are observed  far from the local luminosity-metallicity
relation (i.e., with much lower  metallicities)  need to
have young ages (i.e., be near the start of their evolutionary track) 
and to  follow a  ``closed-box''-like model (see (i) above)  in
order to reach the region of the  luminosity-metallicity
relation of galaxies today, by moving vertically or diagonally and not
horizontally, because they need a substantial increase in [O/H].

\begin{figure}[h!]
\plotone{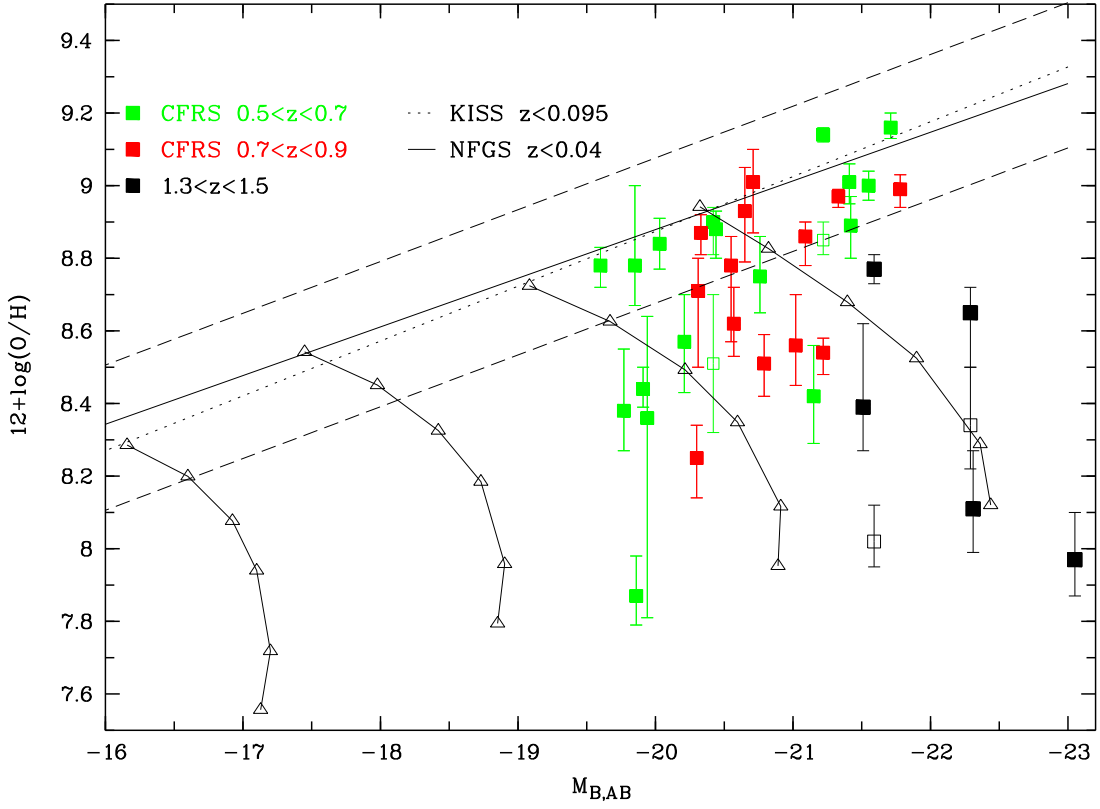}
\caption
{\label{MB_OHcode} \footnotesize Oxygen abundance vs. $M_{B,AB}$ for the $z
  \sim 1.4$ galaxies of our sample (black filled squares). Plotted are also the CFRS
  galaxies, split in two redshift bins: CFRS galaxies at $0.5<z<0.7$
  are plotted as  green filled
  squares, while  CFRS galaxies at $0.7<z<0.9$ are plotted as
  red filled squares.
Open squares are the alternative (but less probable) oxygen abundance
solutions, as discussed in Section 3.3 and Fig.\,4 of Paper\,III. 
 The nearby NFGS and KISS galaxies have [O/H] lying between the dashed lines
  (with the solid and dotted lines indicating the mean local relation
  for NFGS and KISS galaxies, respectively).
 All oxygen  abundances were uniformly measured using the code based on KD02 models.  The
  tracks show a subset of theoretical P\'egase2 models which are consistent
  with the local metallicity-luminosity relation (see text).
}
\end{figure}

~~~ In the following, we
adopt a representative set of  P\'egase2 models to discuss the chemical evolution of
galaxies from $z\sim1.4$ to today. Specifically, we will use models with fixed
$t_1=8$\,Gyrs and $t_{infall}=1$\,Gyr, and different $M_{tot}$ and $p_2$.
These models are highlighted in Fig.\,\ref{MB_OHPegas} by gray/red
lines and filled symbols, and reported in Fig.\,\ref{MB_OHcode}.
  The selected models have a range of parameters from $M_{tot}=2.5\times 10^{11}M_{\odot}$,
$p_2=1$ (right in Fig.\,\ref{MB_OHcode}), to $M_{tot}=5.8\times 10^{9}M_{\odot}$ and $p_2=0.3$
(left). 

We are not so much interested in precise quantitative arguments, but
rather in understanding the qualitative behaviour.
Given the global model properties discussed above, the specific choice
of models in Figs.\,\ref{MB_OHcode} and \ref{MB_OHcodeCADIS} should not affect the general conclusions
we draw below.



\subsection{Results from the comparison  of P\'egase2 models and observations}

~~~   In order to discuss the evolution of the metal abundance of the star forming gas
in galaxies from $z=0$ to $z\sim1.5$, we compare the P\'egase2 models discussed above with the observed
luminosities and metallicities of the five $z\sim1.4$ galaxies
in Fig.\,\ref{MB_OHcode}. Included in the figure are also the intermediate
redshift CFRS galaxies, and the location of local KISS and NFGS galaxies listed in Section \ref{compsampl}.

~~~ It is quite noticeable that there is an age-redshift
 relation along a given P\'egase2 track.
For example,   despite the large
scatter, the bright, $M_{B,AB}<-19.5$, $z\sim1.4$ galaxies (black filled squares) appear to be
``younger'' 
 than $0.7<z<0.9$ galaxies (red filled squares) in the sense that they
 lie towards the beginning of the luminosity-metallicity track. The latter  appear in turn
to be on average ``younger'' than most $0.5<z<0.7$ galaxies (green filled squares)
 which themselves actually overlap
significantly on the diagram with the metallicity-luminosity relation traced
by nearby galaxies. 
 The tracks of the P\'egase2 models suggest that the
bright star forming $z\sim1.4$ galaxies are likely to evolve into the
population of  less luminous but nonetheless rather massive,
metal-rich galaxies that appear in the $0.5<z<0.9$ galaxy population.
The observation of this age-redshift effect is not surprising, and indeed is reassuring!

~~~Subsequent evolution along the P\'egase2 model tracks allows us to
  speculate about what are specifically the $z=0$ descendants of the
  intermediate-$z$ galaxies that we are studying. On the
[O/H] vs.  $M_{B,AB}$ plane at  redshift zero these appear to be massive disk galaxies. The observed
  $I$-band HST morphologies of our $z \ga 0.5$ galaxies cover a wide range, with
  some galaxies having a very compact appearance, and others being extended
  objects, including regular disks (see Paper\,II). These morphologies are
  expected to describe well the distribution of the star forming gas, as they
  picture the $z\sim1$ galaxies in approximately the rest-frame $B$-band,
  where recent star formation dominates the light emission. We therefore
  suspect that the metal-enriched reservoirs of star forming gas that we are
  studying in this paper are being mostly consumed to build up both the disk and the
  bulge components of spiral galaxies.  This interpretation is in agreement
  with other studies which argue for a substantial build up of stellar mass in
  disks \citep[see, e.g.,][]{lilly98} and bulges 
  \citep[see, e.g.,][]{marci97,marci98,marci99,marci01,marci02} at intermediate epochs.

\begin{figure}[h!]
\plotone{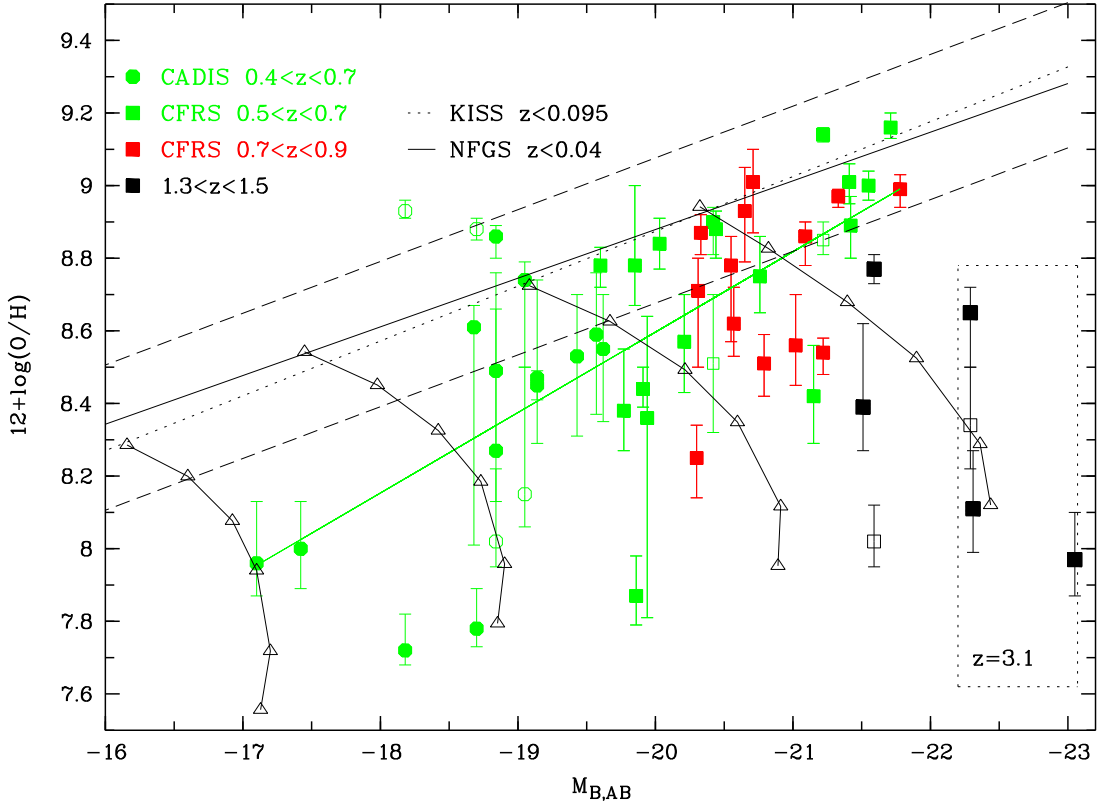}
\caption
{\label{MB_OHcodeCADIS} \footnotesize Same as Fig.\,\ref{MB_OHcode},
  but with additional plotted lower luminosity CADIS
  galaxies (green filled circles), and LBGs galaxies at $z\sim3.1$. 
The location of bright galaxies at $z\sim3.1$ is shown as a box
  encompassing the range of  $\rm{M_{B,AB}}$ and [O/H] (without breaking the \R23\, degeneracy)   derived for these
  objects from Fig. 7 in \citet{pettini01}, which takes also into account the  metallicities
  derived by  \citet{kobko} for $z\sim3.1$
  objects.
Open squares and circles are  alternative (but less probable) oxygen abundance
solutions for some galaxies, as discussed in Section 3.3 and Fig.\,4 of Paper\,III. 
}
\end{figure}

~~~ Adding the low luminosity CADIS galaxies from \citet{maier04} to the metallicity-luminosity diagram
(Fig.\,\ref{MB_OHcodeCADIS}),  we note  an interesting fact:  the rate
of evolution of  galaxies along their tracks appears to depend on
luminosity (mass). What we mean by this is as follows:
 At the  lookback time of about 6\,Gyrs ($z\sim0.6$)
at which a significant fraction of  more massive $M_{B,AB}\la -20$
galaxies are still close to the
zero redshift metallicity-luminosity relation, the lower luminosity (lower
mass) objects at  $M_{B,AB}\ga -19$  sampled by CADIS are already quite
far
from the local metallicity-luminosity.
Moreover, the metallicity-luminosity
relation of the combined CADIS+CFRS $0.4<z<0.7$ sample (green solid
line) shows a change in slope compared to the
local metallicity-luminosity relation, in the sense that the offset in metallicity between
$z\sim0$ and $0.4<z<0.7$ is larger at 
lower luminosity than at higher luminosity.
This may be  due to the fact that lower luminosity (mass) galaxies began
their most rapid evolution 
later than high luminosity (mass) galaxies, as
also suggested by the P\'egase2 models \citep[see also][]{kob03}. 
 As we look back further in time,
we find signs for departures from the $z=0$ metallicity-luminosity
relation occuring at
progessively higher luminosities (or masses).

It is important to address possible selection effects in
Figs.\,\ref{MB_OHcode} and \ref{MB_OHcodeCADIS}. In particular,
there is no observational reason why the $z\sim1.4$ galaxies could not
have had [O/H] $\sim 9$ , nor why the most luminous CFRS objects could not
have had [O/H] $\sim 8.5$. As noted earlier, the CADIS sample may be
somewhat biased towards lower [O/H] by the [OIII]  selection.

To  further clarify the effects seen in Fig.\,\ref{MB_OHcodeCADIS}, one  can see that there are (broadly
speaking) the signatures of the  "downsizing" effect described in Section
\ref{basicobs} in the sense that:
(a) at $z < 0.7$ (green symbols) nearly all galaxies with $M_{B,AB} \la -20$
   are fairly close to the low-z [O/H]-$M_{B}$ relation (there being one
   obvious exception);
(b) at $0.7 < z < 0.9$, this is true only for $M_{B,AB} \la -21.3$; and
(c) at $z \sim 1.4$ even the most luminous galaxies are evolved off of
the low redshift [O/H]-$M_{B,AB}$ relation.

However, it should be noted that the samples of galaxies at  $0.5<z<0.9$  and $z \sim 1.4$
and the range in luminosities (particularly for $z>0.7$) are both quite
small and somewhat heterogeneously selected. 
Therefore, additional measurements of
metallicities at $z>0.5$ of larger samples of galaxies are required to establish a change of the slope of the metalicity-luminosity relation at
these redshifts, as indicated by the analysis of the
metallicity-luminosity relation of our rather small sample.
This could provide   more direct evidence for a    ``downsizing'' scenario.


\section{Conclusions}

The metallicity of the star forming gas has been measured for five galaxies at $z \sim 1.4$ using new near-infrared VLT-ISAAC spectroscopy
in the H- and J-band,
and already available  measurements of the emission line flux of the
\OII\, line, by which the galaxies were selected.
Using the measurements of \OII, \Hb, \OIIIa, \Ha, and upper limits for
\NII\, it was possible to determine the extinction, oxygen abundances,
and extinction corrected star forming rates for these luminous 
($M_{B,AB}<-21.5$) galaxies. The comparison between  P\'egase2  chemical evolution
models, and the observed properties of 
galaxies at $z\sim1.4$, CFRS galaxies at $0.5<z<0.9$,
lower luminosity CADIS galaxies at $0.4<z<0.7$, and 
more luminous Lyman Break Galaxies (LBGs) at $z\sim3.1$, lead us to the following conclusions:

1. The source of gas ionisation in the five $z\sim1.4$  galaxies for
which we measured oxygen abundances  is not
associated with AGN activity, as derived from the log(\OIIIa/\Hb)
versus log(\NII/\Ha) diagnostic diagram. 

2. All the $z\sim1.4$ galaxies have the high \OIIIa/\OII\, and
\OIIIa/\Hb\, line ratios which are typical of the lower luminosity
intermediate-$z$ CADIS galaxies (although it should be noted that CADIS
is [OIII]-selected and may be slightly biased towards high [OIII]/[OII])
and of the more luminous LBGs at higher $z$.
This is in contrast with the
bright, intermediate-$z$ CFRS galaxies, which show low \OIIIa/\OII\, and
\OIIIa/\Hb\, ratios.
The basic trends in our data are consistent
with a ``downsizing'' picture  in the sense that a given type 
of spectrum (high 
\OIIIa/\OII) is found almost exclusively below a luminosity threshold
which is low at zero redshift ($M_{B,AB} \sim -18.5$), increases to 
$M_{B,AB} \sim -20.5$ at $z \sim 0.7$, and has evidently increased to above
$M_{B,AB} \sim -23$ at $z \sim 1.4$.

3. Analysis of a set of P\'egase2 models has shown that  galaxies which are
observed far from the local luminosity-metallicity
relation (i.e., with much lower  metallicities)  need to
have young ages (i.e., be near the start of their evolutionary track) 
and to  follow a  roughly ``closed-box''-like model to reach the region of the  luminosity-metallicity
relation of galaxies today.

4. There is an age-redshift
 relation along a given P\'egase2 track.
The bright, $M_{B,AB}<-19.5$, $z\sim1.4$ galaxies  appear to be
``younger'' 
 than $0.7<z<0.9$ galaxies, and  the latter  appear in turn
to be on average ``younger'' than most $0.5<z<0.7$ galaxies.
 The tracks of the P\'egase2 models suggest that the
bright star forming $z\sim1.4$ galaxies are likely to evolve into the
population of relatively less luminous but nonetheless rather massive,
metal-rich galaxies that appear in the $0.5<z<0.9$ galaxy population.

5. Interestingly,  the rate of evolution of a galaxy along its
track appears to depend on
luminosity (mass).  As we look back  in time,
we find indications for departures from the local metallicity-luminosity relation  occuring at
progessively higher luminosities (or masses). 
This provides further support for a ``downsizing'' picture of galaxy
formation, at least from $z\sim1.4$ to today.


\acknowledgments
We would like to thank the anonymous referee for his or her
suggestions and comments. C.M. acknowledges support
from the Swiss National Science Foundation.


\clearpage
\acknowledgments


%
%

%
%

\clearpage


\clearpage
\clearpage
\clearpage
\clearpage
\clearpage
\clearpage
\clearpage
\clearpage

%
%
\clearpage

\begin{table}[h!]
\caption{\label{ObsISAACz14}  VLT-ISAAC spectroscopic 
  observations of the five $z\sim1.4$ galaxies 
}
\begin{center}
\begin{tabular}{c c c c c c}
\hline
\hline
\#   & Filter & $t_{exp}$(s) & Night & Seeing &  Telluric Standard\tablenotemark{a}\\
\hline
CFRS-221153      & H &  1200 & Oct 03/04 2004& 1\arcsec.0 &  Hip101505 \\
CFRS-221153      & H &  1200 & Oct 06/07 2004& 1\arcsec.0 &  Hip100881 \\
CFRS-221153      & J &  3600 & Oct 04/05 2004& 0\arcsec.8 &  Hip094653 \\
CFRS-221153      & J &  5400 & Oct 06/07 2004& 1\arcsec.2 &  Hip100881 \\
\hline
LTBC-18A         & H &  1800 & Oct 05/06 2004& 1\arcsec.2 &  Hip000183 \\
LTBC-18A         & J &  5400 & Oct 05/06 2004& 1\arcsec.2 &  Hip000183 \\
\hline
LTBC-18B         & H &  2000 & Oct 03/04 2004& 1\arcsec.3 &  Hip109332 \\
LTBC-18B         & H &  1800 & Oct 05/06 2004& 1\arcsec.2 &  Hip000183 \\
LTBC-18B         & J &  6000 & Oct 03/04 2004& 1\arcsec.4 &  Hip109332 \\
LTBC-18B         & J &  5400 & Oct 05/06 2004& 1\arcsec.2 &  Hip000183 \\
\hline
CADIS-23h-3487  & H &  1800 & Oct 06/07 2004& 1\arcsec.0 &  Hip100881\\
CADIS-23h-3487  & J &  5400 & Oct 06/07 2004& 1\arcsec.0 &  Hip100881\\
\hline
CADIS-01h-11134 & H &  1200 & Oct 08/09 2004 & 0\arcsec.9&  Hip108975 \\
CADIS-01h-11134 & J &  1200 & Oct 08/09 2004 & 0\arcsec.9&  Hip108975 \\
\hline
\end{tabular}
\end{center}
\tablenotetext{a}{``Hip'' denotes the
  Hipparcos catalogue from which the telluric standards were selected.
}
\end{table}

\clearpage

\clearpage
\begin{deluxetable}{ccccccccccccc}
\tabletypesize{\footnotesize}
\setlength{\tabcolsep}{0.05in}
\tablewidth{0pt}
\rotate
\tablecaption{\label{VLTISAACz14} \footnotesize Observed and derived quantities for the five
  $z\sim1.4$ galaxies
  }
\tablehead{
\colhead{Nr}    & RA  & DEC &    \colhead{$M_{B,AB}$}   &
\colhead{z} &  \colhead{[OII]\tablenotemark{a}}    &   \colhead{\Hb\tablenotemark{a}}
  &  \colhead{[OIII]\tablenotemark{a}}    &  \colhead{\Ha\tablenotemark{a}}    & \colhead{[NII]\tablenotemark{a,b}}     &
  \colhead{A$_{\rm V}$}
  &     \colhead{SFR}&
  \colhead{[O/H]}
}
\startdata
CFRS-221153  & 22 17 40.60 & +00 18 21.82 & -22.31 & 1.313 & 20$\pm2.0$    &11$\pm1.0$    &30$\pm3.0$    & 34$\pm3.0$ &$<3$   &$0.26^{+0.41}_{-0.26}$&33.04$^{+15.68}_{-8.35}$  &$8.11^{+0.16}_{-0.12}$ \\   
LTBC-18A     & 22 17 40.29 & +00 20 20.60 & -21.59 & 1.445 & 4.3$\pm0.4$   &4$\pm0.8$     &14$\pm1.4$    & 13$\pm1.2$ &$<1.5$ &$0.23^{+0.53}_{-0.23}$&15.67$^{+ 9.51}_{-3.71}$  &$(8.77)^{+0.04}_{-0.10}$\tablenotemark{d}  \\
LTBC-18B     & 22 17 40.15 & +00 20 19.10 & -21.51 & 1.446 & 7.4$\pm0.7$   &3$\pm0.6$     &9 $\pm0.9$    & 7 $\pm0.7$ &$<0.7$ &$0.00^{+0.12}_{-0.00}$& 7.17$^{+ 1.42}_{-0.72}$  &$8.39^{+0.23}_{-0.12}$  \\   
CADIS-23h-3487  & 23 15 28.63 & +11 23 13.54 & -22.29 & 1.440 & 12$\pm1.2$    &7$\pm0.7$     &27$\pm2.5$    & 17$\pm1.5$ &$<2$   &$0.00^{+0.16}_{-0.00}$&17.23$^{+ 4.02}_{-1.72}$  &$(8.65)^{+0.07}_{-0.15}$\tablenotemark{d}  \\           
CADIS-01h-11134 & 01 47 33.37 & +02 22 12.90 & -23.05 & 1.470 & 15$\pm1.5$    &(14$\pm1.4$)\tablenotemark{c}            &32$\pm3.2$    & 39$\pm3.5$ &$<3$   &$0.39^{+0.43}_{-0.39}$\tablenotemark{c}    & 54.98$^{+27.27}_{-17.54}$\tablenotemark{c}  & $7.97^{+0.13}_{-0.10}$\tablenotemark{c}   \\ 
\enddata
\tablenotetext{a}{Fluxes are given in $10^{-17}\rm{ergs}\,\rm{s}^{-1}\rm{cm}^{-2}$}
\tablenotetext{b}{For \NII\, $2\sigma$
  upper limits for the emission line flux are given}
\tablenotetext{c}{\Hb\, of this object coincides with
a strong OH sky line, so the respective flux could not be
measured. However, we derived the \Hb\, flux for this galaxy assuming
$A_{V}=0.5$ and case B recombination, as discussed in the text.}
\tablenotetext{d}{Alternative (but less probable) oxygen abundances found using the method described in
Section 3.3 and Fig.\,4 of Paper\,III are [O/H]=$8.02^{+0.10}_{-0.07}$
for object LTBC-18A, and [O/H]=$8.34^{+0.16}_{-0.12}$ for object  CADIS-23h-3487.
}
\end{deluxetable}



\begin{thebibliography}{}


\bibitem[Carollo et al.(1997)]{marci97}Carollo, C.M., Stiavelli, M., de Zeeuw, P.T., Mack, J., 1997, AJ 114, 2366

\bibitem[Carollo et al.(1998)]{marci98} Carollo, C.M., Stiavelli, M., Mack, J., 1998, AJ 116, 68

\bibitem[Carollo(1999)]{marci99} Carollo, C.M., 1999,  ApJ, 523, 566 


  
\bibitem[Carollo and Lilly(2001, hereafter Paper\,I)]{calilly01} Carollo, C.
  M. \& Lilly, S. J. 2001, ApJ, 548, 153, Paper\,I

\bibitem[Carollo et al.(2001)]{marci01} Carollo, C.M., Stiavelli, M., de Zeeuw, P.T., Seigar, M., Dejonghe, H., 2001, ApJ 546, 2165 

\bibitem[Carollo et al.(2002)]{marci02} Carollo, C.M., Stiavelli, M., Seigar, M., de Zeeuw,
  P.T., Dejonghe, H.,  2002, AJ 123, 159 

\bibitem[Chary \& Elbaz(2001)]{charelbaz01} Chary \& Elbaz, 2001, ApJ, 556, 562

\bibitem[Cowie et al.(1995)]{cowie95} Cowie et al., 1995, Nature 377, 603

\bibitem[Cowie et al.(1996)]{cowie96} Cowie, L. L., Songaila, A., Hu,
  E. M \& Cohen, J. G., 1996, AJ, 112, 839

\bibitem[Crampton \& Lilly(1999)]{cramlilly99} Crampton, D., \& Lilly,
 S.J. 1999, in ASP Conf. Ser. 191, 229 


\bibitem[van Dokkum et al.(2005)]{dokkum05} van Dokkum, P.G., Kriek,
  M., Rodgers, B. et al. 2005, ApJ, 622, 13

\bibitem[Ellison \& Kewley(2005)]{elkew} Ellison, S.L. \& Kewley, L. J., 2005, astro-ph/0508627


\bibitem[Fioc \& Rocca-Volmerange(1999)]{fiocrocca99} Fioc, M. \& Rocca-Volmerange, B., 1999, astro-ph/9912179

\bibitem[Grevesse et al.(1996)]{grevesse} Grevesse, N., Noels, A.,
  Sauval, A. J. 1996. Standard Abundances, ed.
S. S. Holt and G. Sonneborn, Cosmic Abundances:
Proc. of the 6th annual October Astrophysics Conference,
 ASP Conference Ser., 99, 117 (San Francisco: Astron. Soc. of the Pacific)



\bibitem[Hippelein et al.(2003)]{hippe03} Hippelein, H., Maier, C., Meisenheimer, K.  et
 al.  2003, A\&A, 402, 65




\bibitem[Jansen et al.(2000)]{jansen} Jansen, R.~A., Fabricant, D., Franx, M. et al,
  2000, ApJS, 126, 331


\bibitem[Kennicutt(1998)]{ken98} Kennicutt, R. C., Jr. 1998, ARA\&A, 36, 189

\bibitem[Kewley et al.(2001)]{kewley01} Kewley, L.J., Heisler, C.A.,  Dopita, M.A.\&
  Lumsden, S.  2001, ApJS, 132, 37

\bibitem[Kewley \& Dopita(2002, hereafter KD02)]{kewdop02} Kewley, L.J. \& Dopita, M.A. 2002 ApJSS, 142,
  35, KD02

\bibitem[Kobulnicky \& Koo(2000)]{kobko}  Kobulnicky, H. A. \& Koo, D. C. 2000, ApJ, 545, 712


\bibitem[Kobulnicky et al.(2003)]{kob03} Kobulnicky, H. A., Wilmer C. N. A., Weiner, B. J. et al.  2003, ApJ, 599, 1006

\bibitem[Kobulnicky \& Kewley(2004)]{kobkew04} Kobulnicky, H. A. \&
  Kewley, L.J.  2004, ApJ, 617, 240

\bibitem[Le Fevre et al.(1995)]{lef95} Le Fevre, O., Crampton, D., Lilly, S. J., Hammer, F.,
\& Tresse, L., 1995, ApJ, 455, 60


\bibitem[Liang et al.(2004)]{liang04} Liang, Y. C., Hammer, F., Flores,
  H. et al. 2004, A\&A, 423, 867


\bibitem[Lilly et al.(1995a)]{lilly95} Lilly, S. J., Le Fevre, O.,
  Crampton, D., Hammer, F. \& Tresse, L. 1995 ApJ, 455, 50

\bibitem[Lilly et al.(1995b)]{lilly95II} Lilly, S. J., Hammer, F., Le
  Fevre, O., \& Crampton, D.,  1995b, ApJ, 455, 75


\bibitem[Lilly et al.(1996)]{lilly96} Lilly, S. J., Le Fevre, O.,
  Hammer, F., \& Crampton, D. 1996, ApJ, 460, 1

\bibitem[Lilly et al.(1998)]{lilly98} Lilly, S. J., Schade, D., Ellis,
  R. et al., 1998, ApJ, 500, 75

 
\bibitem[Lilly et al.(2003, hereafter Paper\,II)]{lilly03} Lilly, S.J,
  Carollo, C.M. \& Stockton, A. 2003, ApJ, 597, 730, Paper\,II


\bibitem[Madau et al.(1996)]{madau96} Madau, P., Ferguson, H. C.,
  Dickinson, M. E., et al., 1996, MNRAS, 283, 1388

\bibitem[Maier(2002)]{maier02} Maier, C., 2002, PhD Thesis, Naturwissenschaftlich-Mathematische Gesamtfakult\"at der Universit\"at Heidelberg, Germany

\bibitem[Maier et al.(2003)]{maier03} Maier, C.,  Meisenheimer, K., Thommes, E. et
  al. 2003, A\&A, 402, 79

\bibitem[Maier et al.(2004)]{maier04} Maier, C., Meisenheimer, K., Hippelein, H.,
  2004, A\&A, 418, 475

  
\bibitem[Maier et al.(2005, hereafter Paper\,III)]{maier05} Maier, C., Lilly,
  S., Carollo, C. M., Stockton, A. \& Brodwin. M., 2005, ApJ, 634, 849, Paper\,III

\bibitem[Maier et al.(2005, hereafter Paper\,III)]{maier05} Maier, C., Lilly,
  S., Carollo, C. M., Stockton, A. \& Brodwin. M., 2005, ApJ in press,
  astro-ph/0508239, Paper\,III


\bibitem[Melbourne \& Salzer(2002)]{melbsal} Melbourne, J. \& Salzer, J. J. 2002, AJ, 123, 2302



\bibitem[Pagel et al.(1979)]{pagel79}  Pagel, B. E. J., Edmunds, M. G., Blackwell, D. E. et al. 1979, MNRAS, 189, 95


\bibitem[Perez-Gonzalez et al.(2005)]{pergonz05} Perez-Gonzalez, P. G.,
  Rieke, G. H., Egami, E. et al., 2005,  astro-ph/0505101

\bibitem[Pettini et al.(2001)]{pettini01} Pettini, M., Shapley, A. E., Steidel, C. C. 2001, AJ, 554, 981


\bibitem[Pettini et al.(2002)]{pettini02} Pettini, M., Rix, S.A., Steidel, C.C.  et al. 2002, Ap\&SS, 281, 461


\bibitem[Salzer et al.(2005)]{salzer05} Salzer, J. J., Lee, J.C.,
  Melbourne, J. et al., 2005, ApJ, 624, 661


\bibitem[Shapley et al.(2004)]{shapley04} Shapley, A., Erb, D.~K.,
  Pettini, M. et al.  2004, ApJ, 612, 108



\bibitem[Somerville et al.(2001)]{somerv01} Somerville, R.S.,
  Lemson, G., Sigad, Y. et al.  2001, MNRAS, 320, 504

\bibitem[Tran et al.(2004)]{tran04} Tran, K. H., Lilly, S.J.,
Crampton,   D. and  Brodwin, M. , 2004, ApJ, 612, 89




\bibitem[Woosley \&  Weaver(1995)]{woosw} Woosley, S. E. \& Weaver, T. A. 1995, ApJS, 101, 181




\end{thebibliography}
\end{document}